\documentstyle[preprint,aps]{revtex}
\begin {document}
\title{\bf
Total $^4$He Photoabsorption Cross Section Revisited: \\
Correlated HH versus Effective Interaction HH}

\author{Nir Barnea$^1$, Victor D. Efros$^{2}$, Winfried Leidemann$^{3}$
and  Giuseppina Orlandini$^{3}$}
\address{
$^1$ The Racah institute of Physics, The Hebrew University, 91904, 
         Jerusalem, Israel\\
$^{2}$Russian Research Centre "Kurchatov Institute",Kurchatov Square, 1,
123182 Moscow, Russia\\
$^{3}$Dipartimento di Fisica, Universit\`a di Trento, and\\
Istituto Nazionale di Fisica Nucleare, Gruppo collegato di Trento,
 I-38050 Povo, Italy\\
}

\date{\today}
\maketitle

\begin{abstract}
Two conceptually different hyperspherical harmonics expansions are used
for the calculation of the total $^4$He photoabsorption cross section.
Besides the well known method of CHH the recently introduced effective
interaction approach for the hyperspherical formalism is applied.
Semi-realistic NN potentials are employed and final state interaction is fully
taken into account via the Lorentz integral transform method.
The results show that the effective interaction leads to a very good
convergence, while the correlation method exhibits a less rapid convergence
in the giant dipole resonance region. The rather strong discrepancy with
the experimental photodisintegration cross sections
is confirmed by the present calculations.
\end{abstract}

\vfill\eject

The $^4$He photodisintegration in the giant dipole resonance region is a
particularly interesting reaction. An understanding of this process in terms of
a microscopic calculation is not only a challenge in few-body physics, but
could also lead to a deeper insight in the physics of more complex nuclei.
Because of the obvious difficulties in describing correctly the four-nucleon
dynamics, calculations with realistic interactions are still lacking. In the
most advanced calculation \cite{ELO97} semi-realistic NN potential models were
employed and the complicated four-nucleon final state interaction was
treated exactly applying the Lorentz integral transform (LIT) method
\cite{ELO94}. The semi-realistic potential models lead to rather realistic
results for the total three-nucleon photoabsorption cross section \cite{ELOT00}
and the $^4$He inverse energy weighted sum rule (see below). Hence one might
expect a rather realistic description of the $^4$He giant dipole resonance
cross section, which is situated close to the breakup threshold. However, the
obtained cross sections show a considerably more pronounced giant dipole 
resonance than seen in the experimental results published in the last two 
decades (see \cite{ELO97}). On the other hand the experimental
situation is not yet completely settled. Older photoabsorption data (see 
discussion in Ref. \cite{Ber80}) and a more recent determination of the 
photoabsorption cross section via photon scattering \cite{Wells}
show a stronger giant dipole peak. A new round of experiments presently carried
out at Lund will hopefully help to clarify this unsatisfying situation.

Besides clarification on the experimental side it is also necessary to check
the obtained theoretical result. In a recent calculation of the $^4$He
photoabsorption with the effective interaction hyperspherical harmonics (EIHH)
method \cite{BLO00} small deviations from the above mentioned calculation of
Ref. \cite{ELO97} in a correlated hyperspherical harmonics (CHH) approach
were found. The slightly different results are most probably due to not fully
convergent HH expansions. Therefore it is the aim of the present work to study
the convergence in both cases in greater detail extending the calculations to
higher order terms. On the one hand it will allow establishing the correct
$^4$He total photoabsorption cross section with semi-realistic
NN potential models. On the other hand it will show which is the most efficient
HH approach. This is very important in view of calculations of the $^4$He
photoabsorption with realistic forces.

The rate of convergence of an HH expansion is generally rather slow in nuclear
physics problems. In particular the short range repulsion of the NN interaction
leads to high momentum components in the nuclear wave function which can be
parametrized only by including a very large number of HH basis functions $H_n$. 
The convergence can be improved introducing proper short range two-body 
correlation functions $f(r_{ij})$,
\begin{equation}
H_n(\rho,\Omega) \rightarrow \prod_{i,j} f(r_{ij}) H_n(\rho,\Omega)\,,
\end{equation}
where $\rho$ and $\Omega$ denote hyperradius and hyperangle, while $r_{ij}$ is
the relative distance of particles $i$ and $j$. The function $f(r)$ can be
obtained from the solution of the two-body Schr\"odinger equation, since at
short distances the role of other particles is rather unimportant.
Though such correlations lead
to a considerable improvement \cite{FeE72} one still needs in general a rather  
large number of  HH terms in order to reach convergence. The convergence can be 
improved considerably if one introduces longer range correlations \cite{RKV92}. 
However, these two-body long range correlations are less under control, because
correlations among more particles become more important. In addition, different 
from short range correlations, they change considerably from ground to continuum
states. In this respect long range ground state correlations would not be
appropriate for the description of LIT states $\tilde\Psi$ in electromagnetic
disintegrations of nuclei, since these states contain information about the
continuum states.

A quite different approach is the HH effective interaction method. The two-body
Hilbert space is divided in two subspaces with projection operators $P$ and $Q$
($P+Q=1$, $P \cdot Q=0,$ $dim_P=N_p$). In an HH calculation $P$ and $Q$ spaces 
are realized via the hyperangular quantum number $K$ ($P$ space: all HH states
with $K\le K_{max}$). Applying the Lee-Suzuki similarity transformation
\cite{LeeS} one decouples $P$ and $Q$ space interactions
in such a way that one obtains an effective interaction in the $P$ space,
\begin{equation}
V_P= P \left[ \sum_{i<j} V_{i,j} \right]_{eff} P \,.
\end{equation}
The two-body Hamiltonian with this potential has
exactly the same eigenvalues as the lower $N_p$ eigenvalues
of the bare interaction acting on the full $P+Q$ space. Due to the effective
two-body interaction one yields an enormous improvement of the convergence for
the ground state energies of nuclei with A=3-6 \cite{BLO00}. The introduction
of such two-body effective interactions in few-body calculations was first made
for the harmonic oscillator basis \cite{NaB}. In comparison the HH basis offers
further advantages, e.g., the presence of collective HH coordinates allows 
to construct a medium affected two-body effective interaction which leads to
a better convergence.

Since one can interpret the effective interaction as a kind of momentum
expansion one cannot expect that high momentum components are included in a
correct way in the wave function if one works with a small number of HH
functions. On the other hand one should find a much improved
convergence for observables which contain little information on high momentum
components. Thus it is no surprise that also for the nuclear radii an extremely
good convergence was observed in Ref. \cite{BLO00}. Similar good convergence
results are expected to hold for observables that are governed by not too high
momentum components.

Both methods, the correlation (CHH) and the effective interaction approach
(EIHH), will be used in the following study
of the $\gamma + ^4$He $\rightarrow X$ reaction.

We calculate the nuclear $^4$He total photoabsorption cross section
in the dipole approximation
\begin{equation}
\sigma_{tot}(E_\gamma) = 4 \pi^2(e^2/\hbar c) E_{\gamma} R(E_{\gamma})
\label{sigma}
\end{equation}
with
\begin{equation}
R(E_{\gamma}) = \int df |\langle f| D_z|\Psi_0\rangle|^2
\delta(E_f-E_0-E_{\gamma})\,,
\label{response}
\end{equation}
where $D_z$ is the third component of the nuclear dipole operator $\vec D$,
while $E_0$ and $E_f$ denote the eigenvalues of the nuclear Hamiltonian
for ground and final states, respectively, and $E_\gamma$ is the photon energy.
The dipole approximation is well established in low-energy photonuclear
reactions. Deviations from the exact result are expected to be very small for
the total cross section, since also the important meson exchange current
contribution is implicitly taken into account within the dipole 
approximation (Siegert's theorem).

With the LIT method the cross section is calculated indirectly. In a first step
the LIT of the response function $R(E_{\gamma})$, i.e.
\begin{equation} \label{Phi_srsi}
L(\sigma_R,\sigma_I)= \int dE_{\gamma} {R(E_{\gamma}) \over
                         (E_{\gamma}-\sigma_R)^2 + \sigma_I^2}
                       = \langle \tilde\Psi|\tilde\Psi\rangle\,,
\end{equation}
is determined via the asymptotically vanishing LIT state $\tilde\Psi$, which
fulfills the following boundstate like differential equation
\begin{equation}
(H-E_0-\sigma_R +i\sigma_I)|\tilde\Psi\rangle= D_z|\Psi_0\rangle\,.
\end{equation}
The second step of the method consists in the inversion of $L$ in order to
obtain $R(E_{\gamma})$ (see, e.g. \cite{ELO99}).

Formally, one can rewrite equation (\ref{Phi_srsi}) in the following form
\begin{equation}
  L(\sigma_R,\sigma_I) = \sum_{\nu} 
       \frac{ |\langle \Psi_0 | D_z | \tilde\Psi_{\nu} \rangle |^2}
            { (\tilde E_{\nu}-E_0-\sigma_R)^2 + \sigma_I^2   } \;,
\label{phi2}
\end{equation}
where $|\tilde\Psi_{\nu} \rangle $ ($\tilde E_{\nu}$) are eigenfunctions 
(eigenvalues) of $H$ in a truncated space. They are obtained with the same
boundary conditions as $|\Psi_0\rangle$. Generally speaking, these states 
are either bound states or pseudo-resonance states.
It is clear that the low energy (small $\sigma_R$) behaviour of $L$ is 
dominated by the positions of the lowest eigenvalues $\tilde E_{\nu}$.

In the following we compare results obtained with CHH and EIHH methods for the
$^4$He photodisintegration. Different from the above mentioned calculation of
Ref. \cite{ELO97} correlations are also introduced for the ground state wave
function. Our CHH and EIHH calculations agree very well for ground state energy
(-30.69 MeV (CHH), -30.71 MeV (EIHH)) and rms matter radius (1.421 fm (CHH),
1.422 fm (EIHH)) being a bit different from those of Ref. \cite{ELO97} (29.24
MeV, 1.43 fm). However, there is no significant change of the photoabsorption
cross section due to the more precise bound state. In fact one finds a small
reduction of the response function $R(E_{\gamma})$, but due to the higher
binding energy the decrease is compensated by
the increased value for $E_\gamma$ (see Eq.(\ref{sigma})).

In Fig.1 we show the convergence patterns of the transform $L$ using the 
MT-I/III potential \cite{MT} as NN interaction. For the CHH case, depicted in
Fig. 1a, one observes a very rapid convergence for $K_{max}$ values from 1 to 7
($K_{max}$=7 was adopted in Ref. \cite{ELO97}), but for higher $K_{max}$
the convergence is slowed down considerably and $K_{max}=11$ does not yet lead
to a completely convergent result. The EIHH results of Fig.1b show a much nicer
convergence behavior, particularly at low $\sigma$.
This can be understood in view of the fact that the effective interaction 
method is, as mentioned, a kind of momentum expansion and thus  
brings an enormous acceleration to the convergence of the lowest eigenvalues 
which dominate the low-energy cross section (see Eq.(\ref{phi2})).
Besides the CHH results in Fig.1a we illustrate
the EIHH transform for $K_{max}=11$. One sees that
the results are very similar and studying the convergence patterns one might
expect that the converged CHH result will come quite close to the EIHH result.
In the comparison one should also not forget that small differences might
remain even for the converged results, since both calculations are carried out
in completely different ways. In particular the CHH calculation is numerically
less accurate, since it includes a nine-dimensional Monte Carlo integration.

In Fig.2 we illustrate the results for the total $^4$He photoabsorption cross
section obtained from the inversion of the LITs of Fig.1 with $K_{max}=7,\,9,
\,11$. Again one has a very nice convergence for the EIHH case, while the
CHH results are not yet completely convergent. One sees that for increasing
$K_{max}$ the peak of the CHH cross section is shifted to lower energies in
direction of the EIHH peak.

In Fig.3 we show the EIHH cross section results with MT-I/III and TN
\cite{ELO99} potentials ($K_{max}=11$) in comparison to experimental data.
Though the theoretical results are somewhat different from those of Ref.
\cite{ELO97} the comparison with experiment is not improved. One still observes
a considerably stronger giant dipole peak than that found in
photoabsorption experiments. At lower energies the situation has changed a bit,
since the new results exhibit a stronger deviation from experiment. On the
other hand it is evident that the experimental result of Ref. \cite{Wells},
where the total photoabsorption cross section is extracted from Compton
scattering via dispersion relations, agrees  much better with the theoretical 
results 
similarly to the already mentioned older $^4$He photoabsorption data. 

It will be very interesting to get a further clarification of the experimental
cross section from the new experiments at Lund. Since the semi-realistic 
potential models lead to a rather good result for the $^4$He rms radius, which 
is the dominant ingredient in the sum rule for the inverse energy weighted cross
section, one may think that there is not much space to change the 
theoretical cross section. In case of the photodisintegration of the three-
nucleon systems one finds only a 10 \% reduction of the peak cross section due 
to more realistic NN interactions and three-nucleon forces \cite{ELOT00}. On 
the other hand it would be extremely interesting if the difference between
realistic and semi-realistic interactions for the photonuclear cross section
is much higher in the four-nucleon than in the three-nucleon case. 

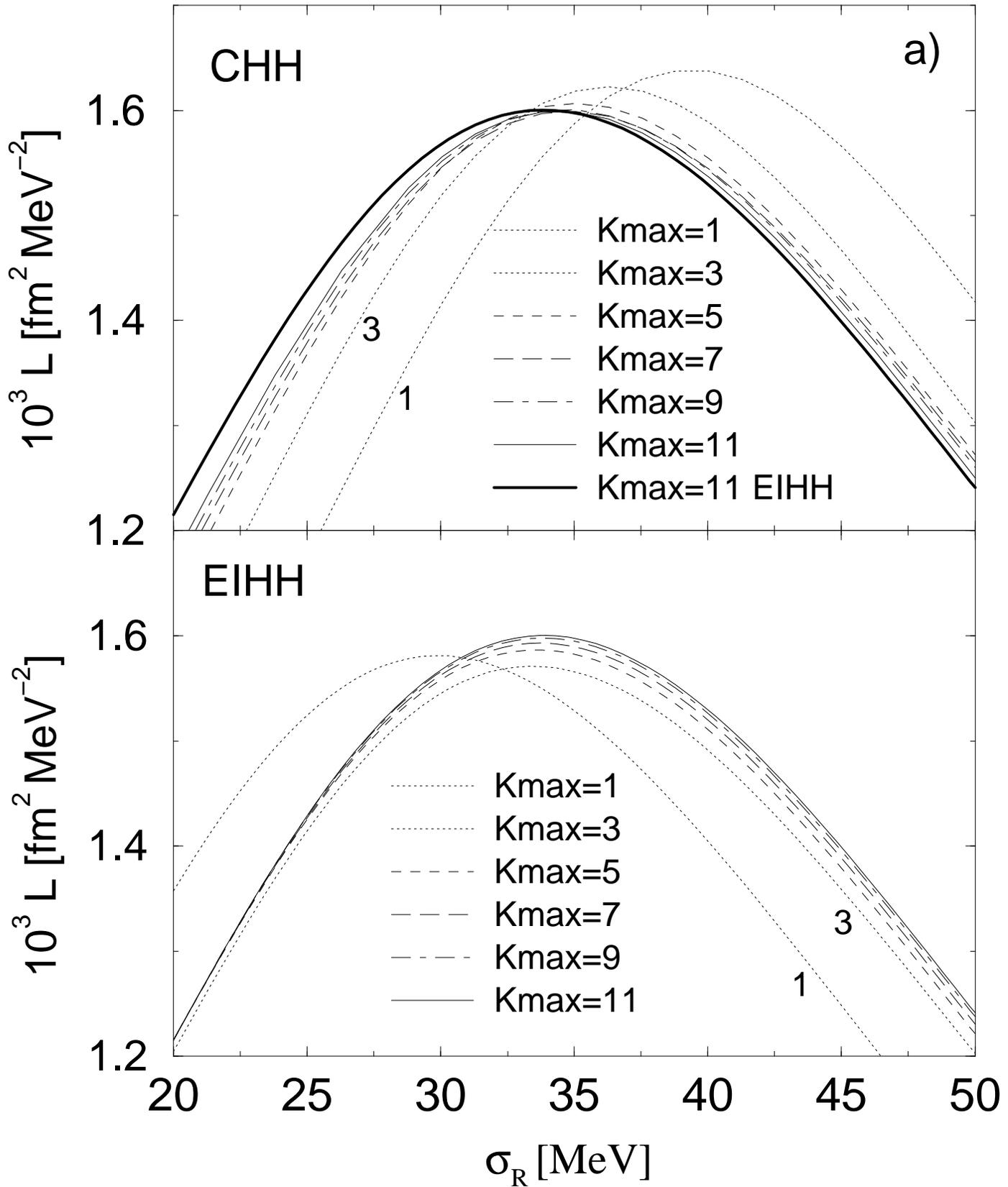
\begin{figure}
\caption{Convergence pattern of $L(\sigma_R,\sigma_I)$ with CHH (a) and EIHH (b)
methods with various maximal
values $K_{max}$ of the hyperangular quantum number $K$ (MT-I/III potential,
$\sigma_I=20$ MeV); in (a) also the EIHH result with $K_{max}=11$ is shown.}
\end{figure}

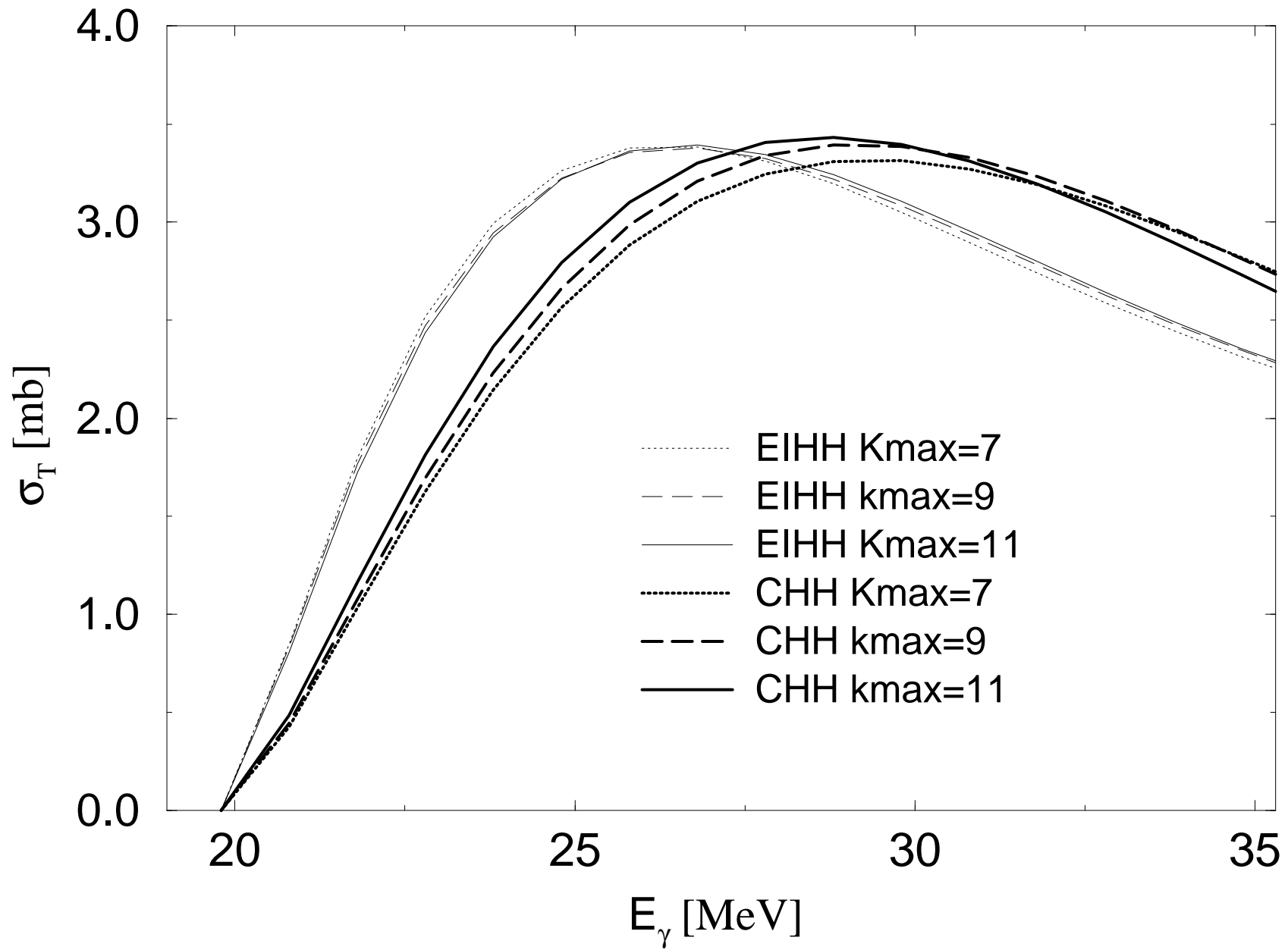
\begin{figure}
\caption{Convergence pattern of the total $^4$He photoabsorption cross section 
for CHH and EIHH methods (MT-I/III potential).}
\end{figure}

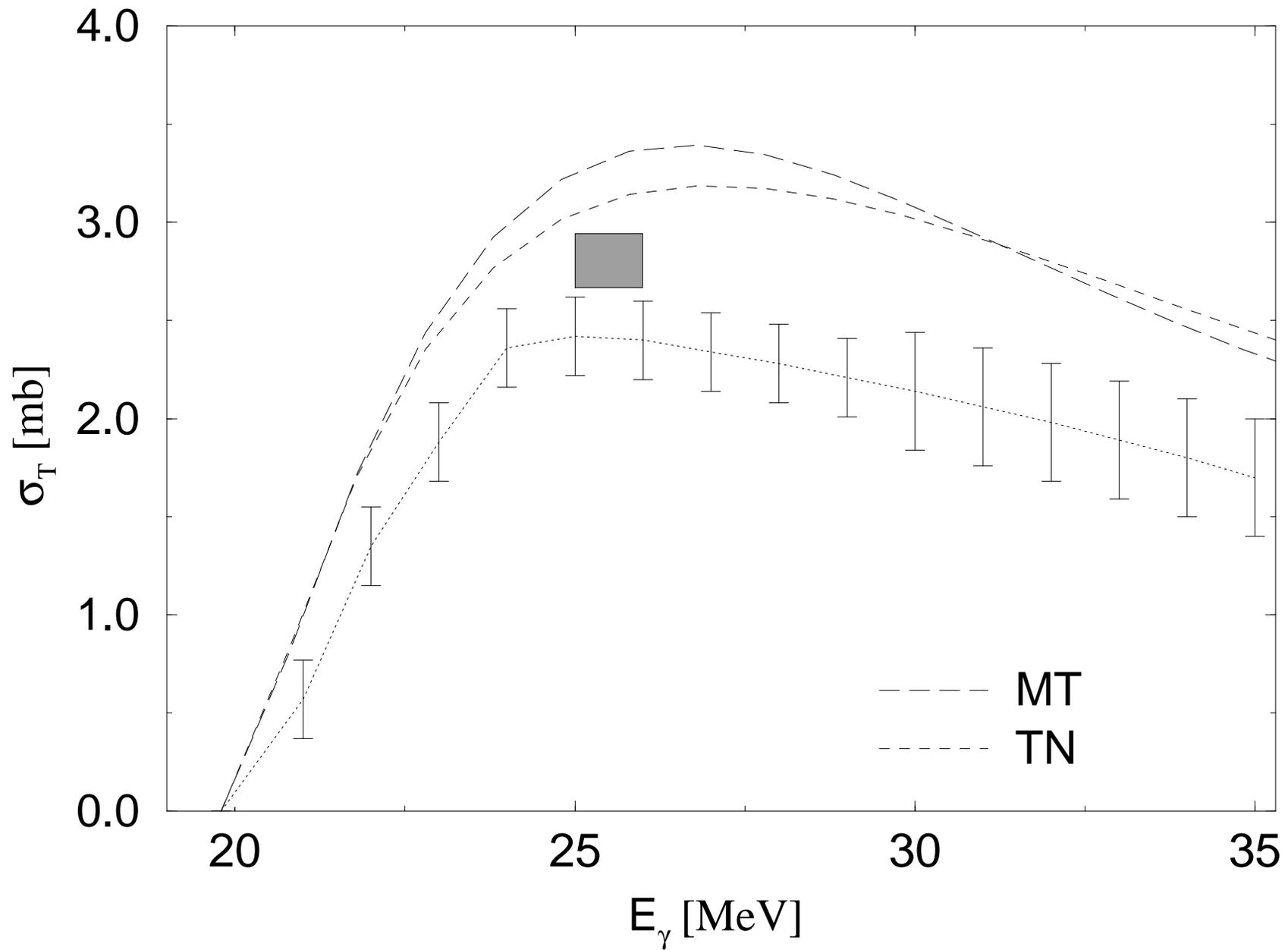
\begin{figure}
\caption{Total $^4$He photoabsorption cross section. Theoretical results
(EIHH method): MT-I/III (long dashed) and TN potentials (short dashed); 
experimental results: sum of $(\gamma,n)$ cross section from [4] and 
$(\gamma,p)^3$H cross section from [13] (dotted curve with error bars) and
indirect determination via Compton scattering from [5] (shaded area).}
\end{figure}

\end{document}